\documentstyle[preprint,aps]{revtex}
\tightenlines

\begin{document}
\newcommand{\beq}{\begin{equation}}
\newcommand{\eeq}{\end{equation}}
\newcommand{\beqa}{\begin{eqnarray}}
\newcommand{\eeqa}{\end{eqnarray}}
\newcommand{\fr}{\frac}
\draft
\preprint{INJE-TP-01-10, hep-th/0112140}

\title{Dynamic dS/CFT correspondence \\
using the brane cosmology}

\author{ Y.S. Myung\footnote{Email-address :
ysmyung@physics.inje.ac.kr}}
\address{Relativity Research Center and School of Computer Aided Science, Inje University,
 Gimhae 621-749, Korea}

\maketitle

\begin{abstract}
We explore the dynamic dS/CFT correspondence using the moving
domain wall(brane) approach in the brane cosmology. The bulk spacetimes
are given by the Schwarzschild-de Sitter (SdS) black hole and the
topological-de Sitter (TdS) solutions. We consider the embeddings of (Euclidean)
moving domain walls into the (Euclidean) de Sitter spaces.
The TdS solution is better to describe the static dS/CFT correspondence
than the SdS black hole, while in the dynamic dS/CFT correspondence
the SdS solution provides situation better than that of the TdS solution.
However, we do not find a desirable cosmological scenario  from
the SdS black hole space.

\end{abstract}
\vfill
Compiled at \today : \number \time.

\newpage

\section{Introduction}
Recently an accelerating universe has proposed to be a way
to interpret the astronomical data of supernova\cite{Per,CDS,Gar}.
Combining this observation with the need of inflation  in the
standard cosmology leads to that our universe approaches de Sitter
geometries in both the far past and the far future\cite{Wit,HKS,FKMP}. Hence it is
very important to study the nature of de Sitter (dS) space\cite{BOU} and the
dS/CFT correspondence\cite{STR}.
 However,
there are two  difficulties in studying de Sitter space\cite{SSV} :
First there is no spatial infinity and   global timelike
Killing vector.  Thus it is not easy to define the conserved  quantities including  mass,
charge and angular momentum appeared in asymptotically  de Sitter space.
Second the dS solution is absent in string theories and thus we
do not   have a definite example  to test the dS/CFT correspondence.

Authors in\cite{BBM} proposed the prescription to calculate the mass of
gravitational field of asymptotically dS spaces. Especially  they put forward  the mass bound
conjecture: {\it Any asymptotically de Sitter space whose mass
exceeds that of de Sitter space contains a cosmological singularity.}
In order to test this conjecture, Cai, Myung and Zhang\cite{CMZ} have
first introduced a topological de Sitter (TdS) solution that always
gives
us a positive mass ($m$) as well as a naked cosmological singularity at $r=0$.
Actually this solution does not  have any black hole horizon but a cosmological
horizon.

The negative mass is found when calculating the mass in  the
Schwarzschild-de Sitter (SdS) black hole\cite{BBM,MYU,GM}.
 Assuming the dS/CFT correspondence,
this induces in turn the negative energy of the dual CFT.
The Casimir energy is also negative which states that
the dual CFT is not unitary. In addition it is not easy to take into
account both its cosmological and  black hole horizons simultaneously.
All of these are circumvented if one
introduces the TdS solution\cite{CAI}.

To understand the dS/CFT correspondence well, one has to
investigate
its dynamic aspects in the brane world cosmology\cite{KRA}.
Nojiri and Odintsov have first discussed  this issue\cite{NO}. Ogushi considered
this correspondence by using the moving domain wall (MDW) approach
in the SdS black hole background\footnote{In the  AdS black hole
background,
the dynamic dS/CFT correspondence was discussed in ref.\cite{PS}.} \cite{OGU}.
On the other hand, Medved\cite{MED1} showed that  the dynamic  dS/CFT correspondence
may be established  in the TdS background.
The mass $m$ located at $r=0$ provides a CFT-radiation
matter on the brane holographically\cite{VS}, and the thermodynamic relation of the CFT
(the Cardy-Verlinde's formula\cite{VER}) coincides with the Friedmann
equation (the dynamic equation for the brane) when the
brane (MDW) crosses the cosmological horizon of  TdS space.
Authors in\cite{OGU,MED1} used the embeddings of Euclidean MDW into dS spaces
with Minkowski signature.

In this paper we explore  dynamic aspects of the dS/CFT
correspondence using the MDW approach in the  SdS and TdS backgrounds.
For this purpose we introduce all kinds of embeddings of the
branes into dS spaces.
In contrast with the static dS/CFT correspondence,
we obtain  the CFT-radiation matter on the dS brane
 moving on the SdS black hole spacetime.
Also we discuss the relation between  thermodynamic CFT and
Friedmann equation.

The organization of this paper is as follows. In section II we
briefly review the results of the static dS/CFT correspondence for
SdS and TdS spaces. We study the embeddings of the
(Euclidean) MDW into (Euclidean) topological-de Sitter space in section III.
In section IV we investigate the embeddings of (Euclidean) MDW in
to (Euclidean) Schwarzschild-de Sitter black hole space.
In section V the relationship between the CFT-thermodynamic
relations and the Friedmann
equation  will be discussed  in
 SdS and TdS spaces.
Finally we discuss our results in section VI.

\section{de Sitter (black hole) solutions}

We start with the $(n+2)$-dimensional Schwarschild-de Sitter metric in the static
coordinates\cite{SSV,MYU,GM}
\beq
ds_{SdS}^2=-\tilde f(r)  dt^2 +
\fr{1}{\tilde f(r)}dr^2 +r^2
d\Omega^2_n,~~\tilde f(r)=1-\fr{\omega_n m }{r^{n-1}}-\fr{r^2}{\ell^2}
\label{2eq1}
\eeq
where $\omega_n= 16 \pi G_{n+2} /nVol(S^n)$ and
$m$ is a parameter related to the black hole mass and $\ell$
is the curvature radius of de Sitter space. The allowed range of $m$
is $0<m \le m_N$ with the Nariai black hole mass
\beq
m_N=\fr{2 \ell^{n-1}}{\omega_n(n+1)} \Big( \fr{n-1}{n+1} \Big)^{\fr{n-1}{2}}.
\label{2eq1s}
\eeq
Beyond this there exists
a naked singularity. We note that Eq.(\ref{2eq1}) is a solution to
the bulk action

\beq
S_{bulk} = {1 \over 16 \pi G_{n+2}} \int d^{n+2} x \sqrt{-g} \left [
 R -2 \Lambda \right]
\label{2eq2}
\eeq
with
$(n+2)$-dimensional positive cosmological constant
$\Lambda=n(n+1)/2\ell^2$. If $m=0$, the solution Eq.(\ref{2eq1})
represents the pure dS solution with a cosmological horizon at
$r_c=\ell$. Using the prescription in\cite{BBM}, it was found that
the gravitational mass  at the future infinity ${\cal I}^+$ is
given by
\beq
M_4=-m,~~~M_5= \fr{3 \pi\ell}{32 G_5} -m
\label{2eq3}
\eeq
which implies that a pure dS$_4$ space has vanishing mass, while a
pure dS$_5$ space has the mass of $3 \pi\ell/32 G_5$. This means
that the masses of the  SdS black holes are always less than those
of pure dS spaces. Assuming the dS/CFT correspondence,
Eq.(\ref{2eq3}) implies  the negative energy of the dual CFT
$(E=-m)$~\footnote{ Usually we neglect the non-vanishing
mass of pure dS spaces (for example, $3 \pi\ell/32 G_5$ in Eq.(\ref{2eq3}))
in calculating the dual CFT energy.}.
Furthermore the Casimir energy $E_{CA}$ of the CFT appears negative,
which implies that the dual CFT is not unitary\cite{CAI}. Hence it seems that
the SdS
black hole solution is not appropriate for our purpose.

The $(n+2)$-dimensional topological-de Sitter metric  is given by \cite{CMZ}
\beq
ds_{TdS}^2\equiv g_{MN}dx^Mdx^N=-f(r)dt^2 +
\fr{1}{f(r)} dr^2 +r^2
\gamma_{ij}dx^idx^j,~~f(r)=k+\fr{\omega_n m }{r^{n-1}}-\fr{r^2}{\ell^2}
\label{3eq1}
\eeq
where $\omega_n= 16 \pi G_{n+2} /Vol(\Sigma_k)$ and
$m$ is assumed to be a positive constant.  $\gamma_{ij}dx^idx^j$ denotes the line element of an
$n$-dimensional hypersurface $\Sigma_k$ with the constant curvature $n(n-1)k$
and its volume $V(\Sigma_k)$. $\Sigma_k$ is given by spherical ($k=1$), flat ($k=0$),
hyperbolic  $(k=-1)$.  This is also a solution to
the action Eq.(\ref{2eq2}).
The case of $k=1$ with the substitution of $m\to-m$ leads to
the previous SdS black hole solution Eq.(\ref{2eq1}). Because of
the action of $m\to-m$, the black hole horizon is absent here.
Instead there exists a
cosmological singularity at $r=0$ for $n\ge 2$.
Using the prescription in\cite{BBM}, it was found that
the gravitational mass (energy) $M$ which is measured  at the future infinity of
${\cal I}^+$ outside the cosmological horizon $r=r_c$, for $k=0$ case, is given by
\beq
M_4=m,~~~M_5= m
\label{3eq2},
\eeq
while for $k=-1$ case,  this is  given by
\beq
M_4=m,~~~M_5= \fr{3 \pi\ell Vol(\Sigma)}{64 \pi G_5} +m.
\label{3eq3}
\eeq
For $k=1$, we have the same
result as in Eq.(\ref{2eq3}) except replacing $m$ by $-m$.
 This means
that the mass (energy) of the  TdS solution are always positive and  greater than
that
of pure dS space. Hence  this confirms the mass bound conjecture in dS space\cite{CMZ}.

Then we apply the TdS/CFT correspondence for
calculating the thermodynamic quantities of the dual CFT.
We can express the total energy of the CFT
($E=m$) in terms of a cosmological horizon $r_c$
given by the maximal root to
$f(r)=0$ :
\beq
 E=m=\fr{r_c^{n-1}}{\omega_n }(r_c^2/\ell^2-k).
 \label{3eq3d}
 \eeq
For $k=0,\pm1$, $E > 0$ is guaranteed from $\omega_n m/r_c^{n-1}>0$ in $f(r_c)=0$.
  We can associate this cosmological horizon with
the Hawking temperature ($T_{TdS}$) and the
entropy $(S_{TdS})$ as
\beq
T_{TdS}= \fr{1}{4 \pi r_c} \Big[ (n+1) \fr{r_c^2}{\ell^2}-(n-1)k
\Big],~~ S_{TdS}=\fr{V_c}{4 G_{n+2}}
\label{3eq4}
\eeq
with the area of the cosmological horizon $V_c=r_c^n Vol(\Sigma)$
in $(n+2)$-dimensional asymptotically dS space. This
corresponds to the volume of the $(n+1)$-dimensional boundary space.
The CFT Casimir energy of $E_{CA}=n(E+ pV-T S)$ with $p=E/nV$ is calculated as
\beq
E_{CA}=- \fr{2n k r_c^{n-1} Vol(\Sigma_k)}{16 \pi G_{n+2}}.
\label{3eq5}
\eeq
For $k=0$ the Casimir energy which is related to the central charge is zero,
 for $k=1$ we have a negative
energy, and for $k=-1$ we have a positive one. Furthermore
the Cardy-Verlinde's formula for $k=\pm1$ is given by\cite{CAI}
\beq
S= \fr{ 2 \pi \ell}{n} \sqrt {|E_{CA}| (2E-E_{CA})},
\label{3eq6}
\eeq
where $S=S_{TdS}$ is the entropy of the cosmological horizon
Eq.(\ref{3eq4}). For $k=0$, one can arrange it as
\beq
S= \fr{ 2 \pi \ell}{n} \sqrt {|E_{CA}/k| (2E-E_{CA})}.
\label{3eq7}
\eeq

Up to now we consider only the static version of the dS/CFT
correspondence. In this case the TdS solution that has a cosmological horizon and a
naked singularity seems to provide a  dS/CFT correspondence
better  than the SdS black hole spacetime.
For a full anlysis of this correspondence  we will discuss the dynamic
evolution of the boundary space in the  TdS background in the next section.

\section{Embedding of MDW into topological-de Sitter space}

In order to define an  embedding of   MDW into the TdS background
Eq.(\ref{3eq1}) properly\footnote{This actually corresponds to a non-trivial $2\to1$
mapping\cite{KRA} : $ r\to a(\tau), t \to t(\tau)$.},
we have to introduce both tangent ($u^M$) and normal ($n_M$) vectors.
This is so because two vectors are essential for defining the projection tensor
of $h_{MN}=g_{MN}-n_Mn_N$, the extrinsic curvature of $K_{MN}=-h_M~^P \nabla_P n_N$
and $K_{\tau\tau}=u^Mu^NK_{MN}$.
First  we usually choose these as in the AdS space
\beq
u^M=(\dot t, \dot a, 0,\cdots,0),~ u^Mu^Ng_{MN}=-1;~~
n_M=(\dot a,  -\dot t , 0,\cdots,0),~ n^Mn^Ng_{MN}=1
\label{3eq7}
\eeq
with $u^Mn_M=0.$
Then  both $u^Mu^Ng_{MN}=-1$ (timelike vector) and $n^Mn^Ng_{MN}=1$ (spacelike vector)
 give us the same
relation
\beq
\fr{1}{f(a)} \dot a^2 - f(a) \dot t^2 =-1
\label{3eq8}
\eeq
which leads to a timelike  brane. Here $f(a)= k+\fr{\omega_n m }{a^{n-1}}
-\fr{a^2}{\ell^2}$. In addition, for a well-defined embedding, we have to
consider the small black hole which satisfies the condition of $\fr{\omega_n m}{\ell^2} <<1.$
Substituting Eq.(\ref{3eq8}) into
the  TdS solution  Eq.(\ref{3eq1}), one has the induced brane metric which takes
 Friedmann-Robertson-Walker (FRW) form
\beq
ds_{TdS}^2 \to ds_{FRW}^2 \equiv h_{\mu\nu}dx^{\mu}dx^{\nu}=
 -d\tau^2 +a(\tau)^2 \gamma_{ij}dx^idx^j,
\label{3eq9}
\eeq
where $h_{\mu\nu}$ is the induced metric on the brane and
the Greek indices $\mu,\nu, \cdots$ denote for the brane
coordinates only.
From the Israel junction condition $K_{\mu\nu}=-\fr{8 \pi G_{n+2}\sigma}{ n} h_{\mu\nu}$
with the brane tension
($\sigma$) together with Eq.(\ref{3eq8}), we obtain
the evolution equation for one-sided brane world scenario\cite{MYU1}
\beq
\dot t= \fr{ 8 \pi G_{n+2} \sigma a}{ n f(a)}=\fr{\tilde \sigma a}{  f(a)},
\label{3eq10}
\eeq
where a reduced tension $\tilde \sigma=8 \pi G_{n+2} \sigma/n$
is introduced for convenience. This
 leads to the first Friedmann equation with $H \equiv \dot a/a$
\beq
H^2=-\fr{f(a)}{a^2}+ \tilde \sigma^2=
  -\fr{k}{a^2} - \fr{\omega_n m}{a^{n+1}} +\fr{1}{\ell^2} +
\tilde \sigma^2,
\label{3eq11}
\eeq
where the dot in ($\dot a$) denotes the differentiation with respect to
the proper time $\tau$.
Here we cannot make any fine-tuning to obtain a flat brane.
We have thus  an effectively   de Sitter brane.
Unfortunately we get a negative energy  from the  cosmological
singularity $(m\not=0)$. This contrasts to the result of its
static TdS/CFT correspondence which says that one gets a positive
energy on the boundary as is shown in Eq.(\ref{3eq3d}).
In the case of finding the negative
energy  in a cosmological model, one wishes  to discard the corresponding
model. Hence we
have to move another case.

Accordingly
 authors in\cite{MED1} used a relation of $ \dot a^2/f(a) - f(a) \dot t^2 =1$
 to obtain
\beq
ds_{TdS}^2 \to ds_{EFRW}^2= d\tau_E^2 +a(\tau_E)^2 \gamma_{ij}dx^idx^j.
\label{3eq11m}
\eeq
Here $\tau_E$ is the Euclidean time obtained by Wick-rotation of $\tau\to
i\tau_E$. Hence the corresponding equation is given by
\beq
H_E^2=\fr{f(a)}{a^2}+ \tilde \sigma^2=
  \fr{k}{a^2} + \fr{\omega_n m}{a^{n+1}} -\fr{1}{\ell^2} +
\tilde \sigma^2
\label{3eq11s}
\eeq
which reduces to with $\tilde \sigma^2=1/\ell^2$ a spacelike flat brane
\beq
H_E^2=\fr{k}{a^2} + \fr{\omega_n m}{a^{n+1}}
\label{3eq11ss}.
\eeq
Here $H_E$ is the Hubble parameter with respect to $\tau_E$ and
thus one finds $H^2_E=-H^2$.
This mapping  from   TdS space with Lorentzian  signature to the spacelike brane with
Euclidean
signature can be  defined properly if one chooses
$u^Mu^Ng_{MN}=1 \to  \dot a^2/f(a) - f(a) \dot t^2 =1$;
$n^Mn^Ng_{MN}=-1 \to  \dot a^2/f(a) - f(a) \dot t^2 =1$
which  is the reversed choice to Eq.(\ref{3eq7}).
For $k=-1,n=3$ case, the spacelike brane starting at $a=0$ crosses the
cosmological horizon $a_c=\sqrt{\omega_3 m -(\omega_3m)^2/\ell^2 }$ of  TdS space
and then reaches the maximum distance $a_m= \sqrt{\omega_3 m}$. And
then it  contracts and crosses the cosmological horizon and finally collapses into
$a=0$. It seems that all cosmological implications  that are derived from AdS space
may be applied to TdS space\cite{MED1}. However, considering Eq.(\ref{3eq11ss})
with $H^2_E=-H^2$, one finds   the negative energy.
Hence it is clear that the dynamic TdS/ECFT
correspondence cannot be  realized in topological-de Sitter space.

The last  case is the embedding of a spacelike brane into  Euclidean TdS space
 by introducing the two spacelike vectors\footnote{ Our convention
 for $n_M$ is just a negative of that in\cite{PAD} because we use
 the extrinsic curvature of $K_{MN}=-h_M~^P \nabla_P n_N$.}
\beq
u^M=(\dot t, \dot a, 0,\cdots,0),~ u^Mu^Ng^E_{MN}=1;~~
n_M=(\dot a, -\dot t , 0,\cdots,0),~ n^Mn^Ng^E_{MN}=1
\label{3eq12}
\eeq
with $g^E_{MN}=$diag$(f,1/f,\cdots)$ and $u^Mn_M=0.$
  Then we have a
relation
\beq
\fr{1}{f(a)} \dot a^2 + f(a) \dot t^2 =1,
\label{4eq1}
\eeq
where the dot denotes differentiation with respect to Euclidean
proper time $\tau_E$. Substituting Eq.(\ref{4eq1}) into
 Euclidean TdS space of Eq.(\ref{3eq1}) leads to the induced brane metric which takes
Euclidean Friedmann-Robertson-Walker (EFRW) form
\beq
ds_{ETdS}^2 \to ds_{EFRW}^2= d\tau_E^2 +a(\tau_E)^2 \gamma_{ij}dx^idx^j.
\label{4eq2}
\eeq
 Considering both Eqs.(\ref{3eq10}) and
(\ref{4eq1}) leads to
\beq
H_E^2=\fr{f(a)}{a^2}- \tilde \sigma^2=
 \fr{k}{a^2} + \fr{\omega_n m}{a^{n+1}} -\fr{1}{\ell^2} -
\tilde \sigma^2.
\label{4eq4}
\eeq
Considering the above equation together with $H^2_E=-H^2$, it is found that this is
equivalent to  Eq.(\ref{3eq11}). In other words, Eq.(\ref{4eq4})
is just a Euclidean version of Eq.(\ref{3eq11}) with a negative energy term.
Let us explore its dynamic ETdS/ECFT correspondence.
We note that  the asymptotic form of the ETdS metric is naively given by
\beq
\lim_{a \to \infty} \Big[\fr{\ell^2}{a^2} ds^2_{ETdS} \Big]= dt_E^2
+ \ell^2 \gamma_{ij}dx^idx^j,
\label{4eq5}
\eeq
which can be identified with the boundary ECFT metric\cite{MAL}. Then the
thermodynamic relations between  the boundary ECFT  and  the bulk ETdS  are given by
\footnote{ For a curved brane, in general, one has to introduce
the relations\cite{PAD,MED2} : $E_{ECFT}={\cal C} m, ~T_{ECFT}= {\cal C}T_{TdS}.$
For a flat brane, one finds ${\cal C}=\fr{\ell}{a}$, whereas for a
curved brane this is given by ${\cal C}=\fr{1}{\tilde \sigma a}$.
Considering $ G_{n+2}= \fr{1}{(n-1)\tilde \sigma} G_{n+1} $, even for the curved brane we
also leads to the same equation as in Eq.(\ref{4eq7}).}

\beq
E_{ECFT}=\fr{\ell m}{a}, ~T_{ECFT}= \fr{\ell
T_{TdS}}{a}, ~~S_{ECFT}= S_{TdS}.
\label{4eq6}
\eeq
Introducing  the energy density $\rho_{ECFT}= E_{ECFT}/V,~ V=a^n Vol(\Sigma)$ and
the pressure $p_{ECFT}=\rho_{ECFT}/n$~\cite{VERS}, Eq.(\ref{4eq4}) can be expressed as the
first Friedmann equation
\beq
H_E^2= \fr{k}{a^2} + \fr{16 \pi G_{n+1}}{n(n-1)} \rho_{ECFT} + \fr{2}{n(n+1)}\Lambda^-_{n+1}
\label{4eq7}
\eeq
with the  cosmological constant $\Lambda^-_{n+1}= -\fr{n(n+1)}{2}(1/\ell^2
+\tilde \sigma^2)$. Here we used the relation between the bulk
and brane newtonian constants for  one-sided brane world
scenario : $ G_{n+2}= \fr{\ell}{n-1} G_{n+1} $.
Eq.(\ref{4eq7}) may imply that the cosmological evolution can be attributed
partly  to
the energy density  of the CFT-radiation matter
originated from the cosmological singularity. But the dynamic
ETdS/ECFT correspondence is not  realized here because one finds a negative
CFT-radiation matter from Eq.(\ref{4eq7})
which comes from the negative energy term in Eq.(\ref{4eq4}). Furthermore, the important
correspondence that
 the Cardy-Verlinde's
formula coincides with the Friedmann equation when the brane
crosses the cosmological horizon $a=a_c$ of TdS space is not found for this case.
Here we can check it easily from Eq.(\ref{4eq4}), by noting that
 $H_E$ is not defined properly ($H_E=i\tilde \sigma$, imaginary) at $a=a_c$,
  the maximal root of $f(a)=0$.

\section{Embedding of MDW into Schwarzschild-de Sitter space}

Up to now we discuss the embedding of the MDW into  TdS space.
From this analysis it is found that although the static behavior
of the  TdS/CFT correspondence is better than that of the SdS/CFT
correspondence, its dynamical TdS/CFT correspondence is not
well-defined.
 Hence it is interesting  to study the dynamical SdS/CFT
 correspondence along the previous section.
  We consider for definiteness  only  the small black hole which satisfies the
 condition of $\omega_n m/\ell^2<<1$.
 For the induced brane metric which takes the FRW form for $k=1$
\beq
ds_{SdS}^2 \to ds_{FRW}^2= -d\tau^2 +a(\tau)^2 d\Omega^2_n,
\label{5eq1}
\eeq
we have the first Friedmann equation\cite{OGU}
\beq
H^2= -\fr{\tilde f(a)}{a^2}+\tilde \sigma^2=
-\fr{1}{a^2} + \fr{\omega_n m}{a^{n+1}} +\fr{1}{\ell^2} +
\tilde \sigma^2.
\label{5eq2}
\eeq
Similarly we cannot make any fine-tuning to obtain a flat brane.
 Eq.(\ref{5eq2}) means that we  have a  de Sitter brane
from  SdS space.
We note here  that  a positive  energy
 can be obtained from the  SdS black hole,
  in contrast with the static SdS/CFT correspondence.
This is a  situation better than that of  the TdS solution.
We  wish to explore its dynamic SdS/CFT correspondence.
This runs closely parallel with the curved brane in the AdS/CFT correspondence\cite{PAD,MED2}.
Its Friedmann equation  is given by
\beq
H^2= -\fr{1}{a^2} + \fr{16 \pi G_{n+1}}{n(n-1)} \rho_{CFT} + \fr{2}{n(n+1)}\Lambda^+_{n+1}
\label{5eq3}
\eeq
with the  positive  cosmological constant $\Lambda^+_{n+1}= \fr{n(n+1)}{2}(1/\ell^2
+\tilde \sigma^2)$ and $ G_{n+2}= \fr{1}{(n-1)\tilde \sigma} G_{n+1} $.
Here $\rho_{CFT}=\tilde E/V= m/\tilde \sigma aV>0$ with $p_{CFT}=\rho_{CFT}
/n$ denotes the CFT-radiation matter which comes from  the SdS black hole
through the dynamic SdS/CFT correspondence.
This contrasts to the static SdS/CFT correspondence
which implies a negative  energy for the dual CFT
($E=-m)$,  as is shown in
Eq.(\ref{2eq3}).

In order to investigate the trajectory of the MDW, we have a
conventional form from Eq.(\ref{5eq2}) as
\beq
\fr{1}{2} \dot a^2 + V(a) =-\fr{1}{2},~~ V(a)= -\fr{\omega_n
m}{2a^{n-1}}- \fr{1}{2} \Big(\fr{1}{\ell^2} + \tilde \sigma^2 \Big)a^2,
\label{5eq4}
\eeq
where  the first term in the left-hand side is the kinetic energy with unit mass and the
second is the potential energy and the term in the right-hand side is the negative total
energy.
The corresponding trajectory  inspired by
Eq.(\ref{5eq4}) can be shown at the Penrose diagram for the SdS black hole\cite{GH}
as was shown in\cite{VS} for the AdS-Schwarzschild black hole case.
The timelike dS brane (MDW) starts at the south pole $a=0$ (Big bang)
and expands with time. At a moment ($\bullet$), the MDW crosses
the black hole horizon $a=a_+$ of SdS space.
For the (3+2)-dimensional small black hole, the location
of the black hole horizon is given  approximately by $a_+=\sqrt{\omega_3 m
+(\omega_3 m)^2/\ell^2}$, while the cosmological horizon is approximated as
 $a_c=\ell \sqrt{1-
\omega_3 m/\ell^2}$. From
Eq.(\ref{5eq2}), we have $H^2=\tilde \sigma^2$ at $a=a_+$. This implies that the
Hubble parameter must obey $H=\tilde \sigma$ at $a=a_+$.
The approximate form of the potential
$V(a)$ is a negative convex ($\cap$). Then  the MDW
reaches the maximum point $ a_m=\sqrt{ \omega_3 m+
(\omega_3 m)^2(1/\ell^2+ \tilde \sigma^2)}$  which is  determined by
$V(a_m)=-1/2$. But
it never  crosses the cosmological horizon at $a=a_c$ because of $a_c>a_m>a_+
$.
This means that the MDW always stays inside the cosmological horizon.
And then the MDW  contracts and will cross the black hole horizon again.
 At this time ($\bullet$)
the Hubble constant will be negative
($H=-\tilde \sigma$) because it is contracting. Finally it falls into the
north pole at $a=0$ (Big crunch).

As was mentioned in\cite{OGU}, the mapping from the SdS
black hole spacetime with Lorentzian signature into the spacelike
brane with Euclidean signature can be  allowed. So this case
is regarded as a candidate for exploring the
dynamic dS/ECFT correspondence.
Its equation is given by $H_E^2=1/a^2-\omega_n m/a^{n+1}$.
Considering $H_E^2=-H^2$, this brane carries with  a positive
energy density from the mass of  SdS black hole.
Also this  can cross both the cosmological and black hole
horizons at $a=a_c,a_+$ where one finds $H_E=\pm \tilde \sigma$.
Hence its dynamic correspondence is established and further its relation
with  the Cardy-Velinde's formula is well-defined.

In the case of the mapping that leads to
EFRW form
\beq
ds_{ESdS}^2 \to ds_{EFRW}^2= d\tau^2 +a(\tau)^2 d\Omega^2_n,
\label{5eq10}
\eeq
one obtains the first Friedmann equation as
\beq
H_E^2= \fr{\tilde f(a)}{a^2}-\tilde \sigma^2=
\fr{1}{a^2} - \fr{\omega_n m}{a^{n+1}} -\fr{1}{\ell^2} -
\tilde \sigma^2.
\label{5eq11}
\eeq
Considering $H^2_E=-H^2$, this is a Euclidean version of  Eq.(\ref{5eq2}) with the positive
CFT-radiation matter.
Eq.(\ref{5eq11})  is very similar to the Euclidean brane in the Euclidean
AdS-Schwarzschild  background\cite{VS} except replacing
the flat brane by the  AdS brane here.
 It is easily checked that
 $H_E$ is not defined properly ($H_E=i\tilde \sigma$, imaginary) at $a=a_c,a_+$,
  the  roots of $\tilde f(a)=0$. Thus the Euclidean brane
neither crosses the cosmological horizon  nor the black hole
horizon in the Euclidean SdS background.

\section{ CFT-thermodynamic relations and Friedmann equations}

We  rewrite the Friedmann equation of  Eq.(\ref{5eq3}) in terms of the Hubble entropy
$(S_H=(n-1)HV/4G_{n+1})$, the Bekenstein-Hawking energy
 $(E_{BH}=n(n-1)V/8 \pi G_{n+1} a^2)$, and $E_{\Lambda}=\Lambda_{n+1}^+ V/8 \pi G_{n+1}$

 \beq
 S_H= \fr{2 \pi a}{n} \sqrt{ E_{BH}[2(\tilde E +E_{\Lambda})-E_{BH}]}
 \label{6eq1}
 \eeq
 which is called the cosmological Verlinde's formula.
On the other hand, we find its static version (that is, the
Cardy-Verlinde's formula) for the cosmological horizon of  the SdS black hole\cite{CAI}
\beq
 S= \fr{2 \pi \ell}{n} \sqrt{ |E_{CA}|(2E-E_{CA})}.
 \label{6eq2}
 \eeq
Even though  $E_{CA},E<0$, it likes to conjecture a naive correspondence between these
 when  considering the
replacements,

\beq
 S_H \to S,~~ E_{BH} \to E_{CA},~~\tilde E +E_{\Lambda} \to E,
 \label{6eq3}
 \eeq
Here the effect of cosmological constant ($E_{\Lambda}$) appears in the
cosmological formula Eq.(\ref{6eq1}). This difference arises because
 in the static case one
usually neglects the energy of pure dS space which relates to the cosmological
constant via $\ell$, as is shown in Eq.(\ref{2eq3}).

One of the striking results for the dynamic AdS/CFT correspondence
is that the Cardy-Verlinde's formula on the CFT-side coincides with
the Friedmann equation
in cosmology when the flat brane crosses the
horizon $a=a_H$ of the AdS-Schwarzschild black hole\cite{VS}. This means that the
Friedmann equation knows the thermodynamics of the CFT.
Let us understand this result in terms of the entropy bounds.
At this point the temperature $(T_H=- \dot H/2\pi H)$  and
 the entropy density $(s=S_H/V=(n-1)H/4G_{n+1})$ can be
expressed in terms of the Hubble parameter and its derivative
only. One introduces the  $\gamma$-function which relates to the central
charge
\beq
\gamma_n(\tau)= \fr{n(n-1)}{16 \pi G_{n+1}} \fr{a_H^{n-1}}{a^{n-1}}=
\fr{n(n-1)}{16 \pi G_{n+1}} \fr{S_{CA}}{S_{BH}},
\label{6eq4}
\eeq
where  $S_{CA}(S_{BH})$ denote the
Casimir entropy bound  of the CFT
(Bekenstein-Hawking entropy bound of $(n+2)$-dimensional bulk space).
In this case one finds the  bound of $\gamma_n(\tau) \le
\fr{n(n-1)}{16 \pi G_{n+1}}$
because  the Verlinde's entropy bound has been  proposed as $S_{CA} \le S_{BH}$.
When $a=a_H$, the Verlinde's bound is saturated. In other words,
the Casimir entropy bound  equals to the Bekenstein-Hawking
entropy bound when the flat brane crosses the horizon (a holographic
point).

It is  very important to study what happens at the moment  when the MDW
 crosses the cosmological horizon at $a=a_c$ in  TdS space.
 Unfortunately it turns out that
 the MDW with a positive CFT-matter does not cross the cosmological horizon.

In the background of the  small SdS black hole,
 the MDW with the positive CFT-matter never crosses the cosmological horizon.
 But it always stays inside the cosmological horizon and crosses the black
  hole horizon at $a=a_+$ because of $a_+<a_m<a_c$.
We do not introduce here the replacements of Eq.(\ref{6eq3}) for the
cosmological horizon. Furthermore, for the black hole horizon in SdS space,
there does not exist any Cardy-Verlinde's  formula like
Eq.(\ref{6eq2})\cite{CAI}. Hence the relation between
the Cardy-Verlinde's formula and the Friedmann equation
 is not established
for the SdS black hole even if the brane crosses the black hole horizon at $a=a_+$.

\section{Discussion}
 In the static dS/CFT
correspondence, TdS space is better than SdS space because
energies  of the TdS-cosmological horizon ($r=r_c$) are always positive, whereas
 energies  of the SdS-cosmological horizon   are always negative.
 Thus  there is no problem in defining the Cardy-Verlinde's
 formula for the TdS space which is regarded  as one of evidences for realizing
  the static dS/CFT
 correspondence. In the SdS black hole we cannot
 define the thermodynamic quantities for the black hole horizon at
 $r=r_+$ properly and thus cannot obtain the corresponding
 Cardy-Verlinde's formula. This means that in the static case an
 observer
 who is located outside the cosmological
 horizon can extract information about the cosmological horizon of the SdS black hole,
 but one cannot investigate  inside the cosmological horizon to  obtain
 further information about the black hole horizon.

On the other hand,  exploring the dynamic dS/CFT correspondence using the MDW
approach in  SdS and TdS spaces leads to the conclusion  that  SdS space is better than TdS
space. In  TdS space we obtain the negative CFT-radiation
matter from the naked singularity, while
in  SdS space we obtain the positive CFT-radiation
matter from the black hole. In the case of finding the negative
energy density in a cosmological model, we have to discard the corresponding
model. Hence the dS  brane moving on the small SdS background gives us a
rather promising model for realizing the dynamic dS/CFT correspondence.
However, the dS brane always stays inside the
cosmological horizon and thus it never cross the cosmological
horizon $a=a_c$ but it can cross the black hole horizon $a=a_+$.
For the cosmological horizon we  obtain a similarity of Eq.(\ref{6eq3}) between
the Cardy-Verlinde's formula and the Friedmann equation, while for
the black hole horizon we cannot define the  Cardy-Verlinde's
formula itself. Hence even for the dS brane  on  SdS space, one does not say
that the Friedmann equation (dynamic equation for the brane) knows
the thermodynamics of the CFT defined on the dS brane.

Furthermore the dynamic SdS/ECFT correspondence is also allowed
because the  embedding of the spacelike brane with Euclidean
signature into  dS spaces with Lorentzian signature is
defined. This supports the static dS/ECFT correspondence based on
the relation between the isometry group of dS space and the
conformal group of the Euclidean boundary space\cite{STR}.

Finally it suggests that
in order to derive the four-dimensional dS model to both get inflation in
the far past and dS geometry in the far future, one may
start with AdS  space\cite{YOUM}.
In this direction  one can obtain the dS brane from  dS
space if the square of  reduced brane tension ($\tilde
\sigma^2$) is greater than the reduced cosmological constant
($1/\ell^2$) in Eq.(\ref{3eq11s}).

\section*{Acknowledgement}
We thank Rong-Gen Cai and  Hyung Won Lee for helpful discussions.
This work was supported in part by the Brain Korea 21
Program of  Ministry of Education, Project No. D-1123.

\end{document}